\documentclass[aps,pra,superscriptaddress,amsmath,amssymb,preprintnumbers,floatfix,showpacs,12pt]{revtex4}
\usepackage{amssymb}
\usepackage{epsfig}
\usepackage{subfigure}
\usepackage{graphicx}
\begin{document}

\title{
Phase  properties of the superposition of squeezed and displaced
number states}

\author{Faisal A. A. El-Orany\footnote{Permanent address: Suez Canal university, Faculty of
Science, Department of mathematics and computer science, Ismailia,
Egypt.}, J. Pe\v{r}ina  } \affiliation{ Joint Laboratory of Optics
of Palack\'y University and Physical Institute of Academy of
Sciences of Czech Republic, 17.~listopadu~50, 772 07~Olomouc,
Czech Republic. }

\author{M. Sebawe Abdalla}

\affiliation{Mathematics Department, College of Science, King Saud
University, P.O. Box 2455, Riyadh 11451, Saudi Arabia}

\begin{abstract}
The phase  properties of superposition of squeezed and displaced
number states are examined in the framework of Pegg-Barnett
formalism. Moreover, amplitude squeezing and phase squeezing are
discussed.

\end{abstract}
 \maketitle

\section{Introduction}
In classical optics, the concepts of the intensity and  phase of
optical fields have a well-defined meaning. That is the
electromagnetic field ($E$) associated with one mode, $E=A\exp
(i\theta)$, has  a well defined amplitude ($A$) and phase
($\theta$). This is not so simple in quantum optics where the mean
photon number and the phase are represented by non-commuting
operators and consequently they cannot be defined well
simultaneously. In fact, the concept of phase is a controversial
problem from the earlier days of quantum optics \cite{scr,{luks}}.
In general there are three methods of treating this issue
\cite{peg}. The first one considers the phase as a physical
quantity in analogy to position or momentum by representing it
with a linear Hermitian phase operator. The second one involves
c-number variables (real or complex) in phase spaces or  their
associated distribution functions, or ensembles of trajectories.
The third one is the operational phase approach in which the phase
information is inferred from the experimental data by analogy with
the classical analysis of the experiment. Each approach has
advantage and disadvantage points. Indeed, the interest in the
phase properties has been motivated by experimental realization of
optical homodyne tomography \cite{tom} allowing quantum phase mean
values to be calculated from the measured field density matrix.

As is well known squeezed states of light (i.e. the states of light with
reduced fluctuations in one quadrature below
the level associated with the vacuum state \cite{yun1}) have become a subject
of intensive studies owing to their interesting applications in various
devices, e.g. in optical communication systems \cite{[5]}, interferometric
techniques \cite{[6]}, and in an optical waveguide tap \cite{[7]}.
Moreover, there are a number of successful experiments
 producing such light \cite{sq}. These states have phase  sensitive noise
 properties and therefore several works have been devoted to follow
 such properties, e.g. for single-mode squeezed states
 \cite{phas, {schl1}}. On the other hand,
the superposition principle is at the heart of quantum mechanics. It implies
that probability densities of observable quantities usually exhibit
interference effects instead of simply being added. The most significant
example reflecting the power of such principle are the Schr\"{o}dinger cat
states \cite{sch1}, which exhibit various nonclassical properties, even
if the original states are close to the classical states
\cite{buz1}. Based on this principle, a general class of quantum states
 has been introduced as a superposition of displaced and squeezed number
 states \cite{[21],{fas}}.
It is important to point out the basis of this class are the squeezed displaced
 number states \cite{[27]} which are purely nonclassical states.
This class is represented as a single mode vibration of the electromagnetic
field suddenly squeezed-plus-displaced by a collection of two
displacements $180^{0}$ out of phase with respect to each other
\cite{[21],{fas}}, i.e.

\begin{equation}
|r,\alpha ,n\rangle _{\epsilon }=\lambda _{\epsilon
}[\hat{D}%
(\alpha )+\epsilon \hat{D}(-\alpha )]\hat{S}(r)|n\rangle ,\label{1}
\end{equation}

\noindent where $\lambda _{\epsilon }$ is the normalization constant,
$\hat{D}(\alpha )$
and $\hat{S}(r)$ are displacement and squeeze operator respectively, while
$\alpha $ and $r$ are displacement and squeeze parameters; $\epsilon
=|\epsilon |e^{i\phi }$ is a complex parameter, and $|n\rangle $
denotes a
Fock state. Squeeze and displacement operators are given, respectively, by

\begin{equation}
\hat{S}(r)=\exp[ \frac{r}{2}( \hat{a}^{2}- \hat{a}^{\dagger2})],\label{2}
\end{equation}

\begin{equation}
\hat{D}(\alpha )=\exp (\hat{a}^{\dagger}\alpha-\hat{a}\alpha^{*}),\label{3}
\end{equation}

\noindent where $\hat{a}$ and $\hat{a}^{\dagger }$ are annihilation and
creation operators. The normalization constant $\lambda _{\epsilon }$ is
given by

\begin{equation}
| \lambda_{\epsilon }^{2}| ^{-1}= 1+|\epsilon |^{2}+2|\epsilon | \exp
(-2|t|^{2} ) L_{n}(4|t|^2) \cos \phi ,  \label{4}
\end{equation}

\noindent with

\begin{equation}
t=\alpha \cosh r+\alpha^{*}\sinh r,
\label{5}
\end{equation}

\noindent and $L_{n}(.)$ is the Laguerre polynomial.
The density matrix $\hat{\rho}$ of this state can be written as
\begin{equation}
\hat{\rho}=
| \lambda_{\epsilon }|^{2}(\hat{\rho}_{M}+\hat{\rho}_{I}).  \label{501}
\end{equation}
The part of the density matrix corresponding to the statistical mixture of
two squeezed displaced number states is:
\begin{equation}
\hat{\rho}_{M}=
 \hat{D}(\alpha )\hat{S}(r)|n\rangle
\langle n| \hat{S}^{\dagger}(r)\hat{D}^{\dagger}(\alpha )
+|\epsilon|^{2} \hat{D}(-\alpha )\hat{S}(r)|n\rangle
\langle n| \hat{S}^{\dagger}(r)\hat{D}^{\dagger}(-\alpha ),  \label{52}
\end{equation}
while the quantum interference part has the form
\begin{equation}
\hat{\rho}_{I}=
 \epsilon^{*}
 \hat{D}(\alpha )\hat{S}(r)|n\rangle
\langle n| \hat{S}^{\dagger}(r)\hat{D}^{\dagger}(-\alpha )
+\epsilon \hat{D}(-\alpha )\hat{S}(r)|n\rangle
\langle n| \hat{S}^{\dagger}(r)\hat{D}^{\dagger}(\alpha ).  \label{53}
\end{equation}
This quantum interference part of the density matrix contains information
about the quantum interference between component states
$\hat{D}(\pm\alpha )\hat{S}(r)|n\rangle$ and this will be responsible for
some interesting behaviour of the phase distribution, as we will see.

For completeness, the physical interpretation of such states can be related
to a superposition
of coherent states formed due to two excitations on  particularly excited
harmonic oscillators \cite{[27]}.
 It is clear that these states enable us to
obtain generalizations of some results given in the
literature. It has been shown that these states can be
generated, by means of the so-called quantum state engineering, and also by
means of trapping ions (for more details, see refs.\cite{[21],{fas}}).
The quantum properties of these states reveal that they can exhibit
sub-Poissonian statistics, quadrature squeezing and oscillations in
photon-number distribution. Moreover, the influence of thermal noise on
the behaviour of such  superposition has been considered \cite{fais}
showing that the correlation between different oscillators is essentially
responsible for nonclassical effects similar to Schr\"{o}dinger cat states.
This fact has been demonstrated from the behaviour of Wigner function
and photon-number distribution.

In this article we  study the phase properties of the
superposition (\ref{1}) using Pegg-Barnett technique \cite{pegg}
which is most convenient for the current problem.
This investigation is organized as follows:
In section 2 we give the description of the Pegg-Barnett technique
and the basic relations related to the state (\ref{1}),
followed by section 3  where the results
are discussed. The  conclusions are summed in section 4.

\section{Basic relations}

Here we give essential background for Pegg-Barnett \cite{pegg} phase
formalism and the basic relations for the state under discussion.
Their formalism is based on introducing a finite $(s+1)$-dimensional space
$\Psi$ spanned by the number states $|0\rangle,|1\rangle,...,|s\rangle$.
The physical variables (expectation values of Hermitian operators)
are evaluated in the finite dimensional space $\Psi$ and  at
the final stage the limit $s\rightarrow \infty$ is taken.
A complete orthonormal basis of $s+1$ states is defined on $\Psi$ as

\begin{equation}
|\Theta_{m}\rangle =\frac{1}{\sqrt{s+1}}\sum^{s}_{k=0}\exp (ik\Theta_{m})
|k\rangle, \label{6}
\end{equation}
where

\begin{equation}
\Theta_{m}=\Theta_{0}+\frac{2\pi m}{s+1}, \qquad m=0,1,...,s.
\label{7}
\end{equation}
The value of $\Theta_{0}$ is arbitrary and defines a particular basis
of $s+1$ mutually orthogonal states. The Hermitian phase operator is
defined as
\begin{equation}
\hat{\Phi}_{\theta}=\sum^{s}_{m=0}\Theta_{m} |\Theta_{m}\rangle
\langle \Theta_{m}|,
\label{8}
\end{equation}
where the subscript shows the dependence on the choice of $\Theta_{0}$.
The phase states (\ref{6}) are eigenstates of the phase operator (\ref{8})
with the eigenvalues $\Theta_{m}$ restricted to lie within a phase window
between $\Theta_{0}$ and $2\pi+\Theta_{0}$.
The expectation value of the phase operator (\ref{8}) in a pure state
$|\psi\rangle=\sum^{\infty}_{m=0}C_{m}|m\rangle$, where $C_{m}$ is the
 weighting coefficient including the normalization constant,
is given by
\begin{equation}
\langle \psi |\hat{\Phi}_{\theta}|\psi\rangle =\sum^{s}_{m=0}\Theta_{m}
|\langle \psi|\Theta_{m}\rangle|^{2}.
\label{9}
\end{equation}

The density of phase states is $(s+1)/(2\pi)$, so the continuum phase
distribution as $s$ tends to infinity is

\begin{eqnarray}
\begin{array}{rl}
P(\Theta)={\rm lim}_{s\rightarrow \infty}\frac{s+1}{2\pi}|\langle
\Theta_{m}|\psi\rangle|^{2} \\
\\
=\frac{1}{2\pi} \sum_{m,m^{'}=0}^{\infty} C_{m}C^{*}_{m^{'}}
\exp[i(m-m^{'})\Theta],
\end{array} \label{10}
\end{eqnarray}
where $\Theta_{m}$ has been replaced by the continuous phase variable
$\Theta$. As soon as the phase distribution $P(\Theta)$ is known, all
the quantum-mechanical phase moments can be obtained as a classical
integral  over $\Theta$. The phase distribution is normalized
such as
\begin{equation}
\int^{\pi}_{-\pi}P(\Theta)d\Theta=1. \label{11}
\end{equation}

One of the particular interesting quantities in the description of the phase
is the phase variance determined by

\begin{equation}
\langle (\triangle \hat{\Phi})^{2}\rangle =\int \Theta^{2} P(\Theta)d
\Theta -\left(\int \Theta P(\Theta)d\Theta \right)^{2}.\label{12}
\end{equation}
As we mentioned earlier the mean photon number and the phase are conjugate
quantities in this approach and consequently they obey the
following uncertainty relation

\begin{equation}
\langle (\triangle \hat{\Phi})^{2}\rangle
\langle (\triangle \hat{n})^{2}\rangle \geq \frac{1}{4}
|\langle [\hat{n},\hat{\Phi}]\rangle|^{2}. \label{13}
\end{equation}

The number--phase commutator appearing on the right hand side of (\ref{13})
can be calculated for any physical state \cite{pegg} as
\begin{equation}
\langle [\hat{n},\hat{\Phi}]\rangle=i[1-2\pi P(\Theta_{0})].\label{14}
\end{equation}
In relation to (\ref{13}), we can give  the notion of the number and phase
squeezing \cite{wod,{buz2}} through the relation

\begin{equation}
S_{n}=\frac{\langle (\triangle \hat{n})^{2}\rangle}
{\frac{1}{2}|\langle [\hat{n},\hat{\Phi}]\rangle|}-1,\label{15}
\end{equation}

\begin{equation}
S_{\theta}=\frac{\langle (\triangle \hat{\Phi})^{2}\rangle}
{\frac{1}{2}|\langle [\hat{n},\hat{\Phi}]\rangle|}-1.\label{16}
\end{equation}
The values of $-1$ of these equations means maximum squeezing of the
photon number or the phase.

We shall use above relations to study the phase distribution for the
superposition of displaced and squeezed number states (\ref{1}). In this case
$C_{m}=C_{m}(r,\alpha,n,\epsilon)=
\langle m|r,\alpha,n\rangle_{\epsilon}$ and this quantity
can be calculated through the identity
\begin{equation}
\langle  m|n ,\alpha
,r\rangle_{\epsilon}=\int_{-\infty}^{+\infty}dx
\Upsilon^{*}_{m}(x)
\Upsilon^{(\epsilon)}_{n}(x,r,\alpha)
, \label{300}
\end{equation}
where
$\Upsilon^{(\epsilon)}_{n}(x,r,\alpha)=
\langle x|n,r,\alpha\rangle_{\epsilon} $ is the wavefunction of
(\ref{1}) having the form
 \cite{fas}

\begin{eqnarray}
\begin{array}{rl}
\Upsilon^{(\epsilon)}_{n}(x,r,\alpha)=\frac{\lambda_{\epsilon}{\rm e}
^{\frac{r}{2}}}
{\sqrt{2^{n}n!}}(\frac{\omega}{\pi \hbar})^{\frac{1}{4}}\left\{
\exp\left[
-\frac{{\rm e}^{2r}}{2}(x\sqrt{\frac{\omega}{\hbar}}-\sqrt{2}\alpha)^{2}
\right] \right.
\\
\\
\times {\rm H}_{n}[{\rm e}^{r}(x\sqrt{\frac{\omega}{\hbar}}-\sqrt{2}\alpha)]
 +\epsilon
\exp\left[ -\frac{{\rm e}^{2r}}{2}(x\sqrt{\frac{\omega}{\hbar}}+\sqrt{2}\alpha)^{2}
\right]
\\
\\
\times \left. {\rm H}_{n}[{\rm e}^{r}(x\sqrt{\frac{\omega}{\hbar}}+\sqrt{2}\alpha)]\right\},
\end{array}
\label{301}
\end{eqnarray}
where $\alpha$ is real, $H_{n}(.)$ is the Hermite polynomial of order $n$;
$\omega$ and $\hbar$ are frequency of the harmonic oscillator and
 Planck's constant divided by $2\pi$.
Substituting from (\ref{301}) together with the wavefunction $\Upsilon_m(x)$
of the Fock state (which can be deduced from (\ref{301}) by simply setting
$r=\alpha=\epsilon=0$) into (\ref{300}) and carrying out the integration,
using  the identity \cite{[50]}

\begin{equation}
\begin{array}{rl}
\sqrt{\frac{M}{\pi}}\int_{-\infty}^{+\infty}dx
{\rm H}_{m}(x){\rm H}_{n}(\Lambda x+d) \exp(-Mx^{2}+cx)=
\exp (\frac{c^2}{4M})
\\
\\
\times (\sqrt{\frac{M-1}{M}})^{m}
 (\sqrt{\frac{M-\Lambda ^{2}}{M}})^{n}
 \sum_{j=0}^{min(m,n)}  \frac{n! m!}
{j!(n-j)!(m-j)!}
\\
\\
\times \left(\frac{2\Lambda}{\sqrt{(M-1)(M-\Lambda^{2})}}\right)^{j}
 {\rm H}_{m-j}\left(\frac{c}{2\sqrt{(M-1)M}}\right)
{\rm H}_{n-j}\left(\frac{c\Lambda+2dM}{2\sqrt{M(M-\Lambda)}}\right),
\end{array}
\label{302}
\end{equation}
we arrive at \cite{fas}

\begin{eqnarray}
\begin{array}{rl}
 \langle  m|n ,\alpha
,r\rangle_{\epsilon}=\frac{
\lambda_{\epsilon }(\frac{\tanh r}{2})^{\frac{m}{2}}}{\sqrt{n!
m!\cosh r}}
\exp [\frac{\tau ^{2}}{2}(\tanh r-1)]
\sum_{j=0}^{min(m,n)}  \frac{n! m!}
{j!(n-j)!(m-j)!}[\frac{2}{\sqrt{\sinh 2r}}]^{j}
\\
\\
\times [\frac{-\tanh r}{2}]^{\frac{(n-j)}{2}}
 {\rm H}_{n-j}(\frac{i\alpha}{\sqrt{\sinh 2r}})
{\rm H}_{m-j}(\frac{\tau}{\sqrt{\sinh 2r}})[1+(-1)^{(n+m)}\epsilon ],
\end{array}
\label{17}
\end{eqnarray}
where  $\tau=\alpha\exp (r)$.
 Another way for  the derivation of
(\ref{17}) can be found
in the appendix of \cite{kra}. When $r\rightarrow 0$,  (\ref{17})
reduces to the distribution
coefficient for the superposition of displaced Fock states
as

\begin{eqnarray}
\begin{array}{rl}
C_{m}(r=0,\alpha,n,\epsilon)=
\lambda_{\epsilon}\sqrt{n!m!}\exp(-\frac{\alpha^{2}}{2})
\sum_{j=0}^{min(m,n)}  \frac{ (-1)^{n-j}\alpha^{n+m-2j}}
{j!(n-j)!(m-j)!}[1+(-1)^{n+m}\epsilon ]
\\
\\
=\lambda_{\epsilon}\left[\frac{p!}{q!}\right]^{\frac{1}{2}}
(-1)^{n-p}\alpha^{n+m-2p}\exp(-\frac{\alpha^2}{2})[1+(-1)^{n+m}\epsilon]
{\rm L}^{q-p}_{p}(\alpha^{2}),
\end{array}
\label{18}
\end{eqnarray}
where  $L^{\gamma}_{n}(.)$ is the associated Laguerre
polynomial, and $p={\rm min}(n,m)$ and $q={\rm max}(n,m)$.  The photon
number variance  is defined as

\begin{eqnarray}
\begin{array}{rl}
\langle (\triangle \hat{n})^{2}\rangle=
\langle \hat{n}^{2}\rangle-
\langle \hat{n}\rangle^{2}\\
\\
=\langle \hat{a}^{\dagger 2}\hat{a}^{ 2}\rangle
+\langle \hat{n}\rangle -\langle \hat{n}\rangle^{2},
\end{array}   \label{19}
\end{eqnarray}
where the number operator $\hat{n}=\hat{a}^{\dagger}\hat{a}$.
In terms of (1)
the quantities $\langle \hat{a}^{\dagger 2}
\hat{a}^{ 2}\rangle$ and $\langle \hat{n}\rangle$
 can be straightforwardly calculated,
however, the explicit forms can be found  in \cite{fas}.
For completeness, the phase variance of (1) reads
\begin{eqnarray}
\begin{array}{rl}
\langle (\triangle \hat{\Phi})^{2}\rangle=\frac{\pi^{2}}{3}+4{\rm Re}
\sum_{m>m'}C_{m}(r,\alpha,n,\epsilon)C^{*}_{m'}(r,\alpha,n,\epsilon)
\frac{(-1)^{m-m'}}{(m-m')^{2}}
\\
\\
  -4\left[{\rm Re} i\sum_{m>m'}C_{m}(r,\alpha,n,\epsilon)C^{*}_{m'}
  (r,\alpha,n,\epsilon) \frac{(-1)^{m-m'}}{(m-m')}
\right]^{2}.
\end{array}\label{20}
\end{eqnarray}
The value $\pi^2/3$ is the phase variance for a state with uniformly
distributed phase, e.g. the vacuum state.
Finally, for the future purpose,
we give the form of Wigner function of (\ref{1}) \cite{[21], {fas}}

\begin{equation}
\begin{array}{rl}
W^{(\epsilon )}(x,y )=\frac{2(-1)^{n}| \lambda_{\epsilon} | ^{2}}{\pi} \left\{
|\epsilon |^{2}\exp [-2(y^{2} {\rm e}^{-2r}+{\rm e}^{2r}(x+\alpha)^{2})] \right.
\\
\\
\times
 {\rm L}_{n}[4(y^{2}
{\rm e}^{-2r}+{\rm e}^{2r}(x+\alpha)^{2})]
+\exp [-2(y^{2} {\rm e}^{-2r}+{\rm e}^{2r}(x-\alpha)^{2})]
\\
\\
\times  {\rm L}_{n}[4(y^{2}
{\rm e}^{-2r}+{\rm e}^{2r}(x-\alpha )^{2})]
+2 \exp [-2(y^{2} {\rm e}^{-2r}+{\rm e}^{2r} x ^{2})]
\\
\\
\times \left. |\epsilon |
{\rm L}_{n}[4(y^{2}
{\rm e}^{-2r}+{\rm e}^{2r} x^{2})]\cos(4y\alpha -\phi )\right\}.
 \label{24}
\end{array}
\end{equation}

Based on the results of the present section we discuss the phase
properties of the state (\ref{1}) in the following section.

\section{Discussion of the results}
For understanding the behaviour of the phase distribution  of the states
 (\ref{1}) it is more
convenient firstly to study such properties for two  subsidiary states which
are the superposition of displaced number states \cite{oba}, and
squeezed and displaced number states \cite{[27]}.
These two states are of interest because it has been shown for the former
states that they can exhibit strong sub-Poissonian character as well as
quadrature squeezing. The  bunching and antibunching properties for the
latter states have been discussed in \cite{mah}.  Investigation
of such properties
for these two kinds of states most clearly illustrates the
role of the different parameters in the state (\ref{1}) in the behaviour
of the distribution.

In the following two subsections we discuss the phase distribution determined
by (\ref{10}), and the phase variance and phase squeezing, respectively.
For simplicity  we restrict our investigation to real
values of $\alpha$ and $\epsilon$.
\subsection{Phase probability distribution}
In general we find that there are two regimes controlling  the behaviour
of the phase distribution  for the states under discussion which are
$\alpha>>1$ and $\alpha \leq 1$ provided that $n$ is finite. For the
first case, the even- and
odd-cases (i.e. $\epsilon=1$ and $-1$) provide  similar behaviours and
there is one-to-one correspondence  between the components of density matrix
(\ref{501}) and that corresponding to the curves.
In other words, $\hat{\rho}_{M}$ is responsible for the central behaviour
(i.e. around $\Theta =0$) of the phase distribution and $\hat{\rho}_{I}$
is responsible for the lateral behaviour, i.e. when $\Theta\rightarrow \pm
\pi$ (see Fig. 2a).
Nevertheless, all these features are washed out
in the second regime and the behaviour  dramatically
change. Such behaviours can be understood well if we turn our attention to
the behaviour of the $W$-function where the interference in phase space
may be seen clearly.
More illustratively,  $W$-function (\ref{24}) includes three terms, the first
two terms representing  the statistical mixture of squeezed displaced
number states and the third one is
the interference part. When $\alpha \leq 1$ (as shown in Fig. 1a and b for
superposition of displaced number states) the contributions of these
components are comparable so that even- and odd-cases
distributions have different shapes.
On the other hand, when $\alpha>>1$ the structures of the $W$-function
for  even- and odd-cases
are almost similar, i.e. they include two symmetrical  peaks originated
 at $\pm\alpha$ and interference fringes are in between.
Of course, such situation is still valid if
squeezing in the displaced superimposed number state in optical cavity is
considered, however, the peaks will be then stretched.
This  behaviour is reflected in the behaviour of the phase
distribution, as we will see.
It should be stressed that the greatest degree of the nonclassical behaviour
for the superposition (1) actually occurs for rather low values of $\alpha$
\cite{[21],{fas}}.

\begin{figure}[h]%
  \centering
  \subfigure[]{\includegraphics[width=8cm]{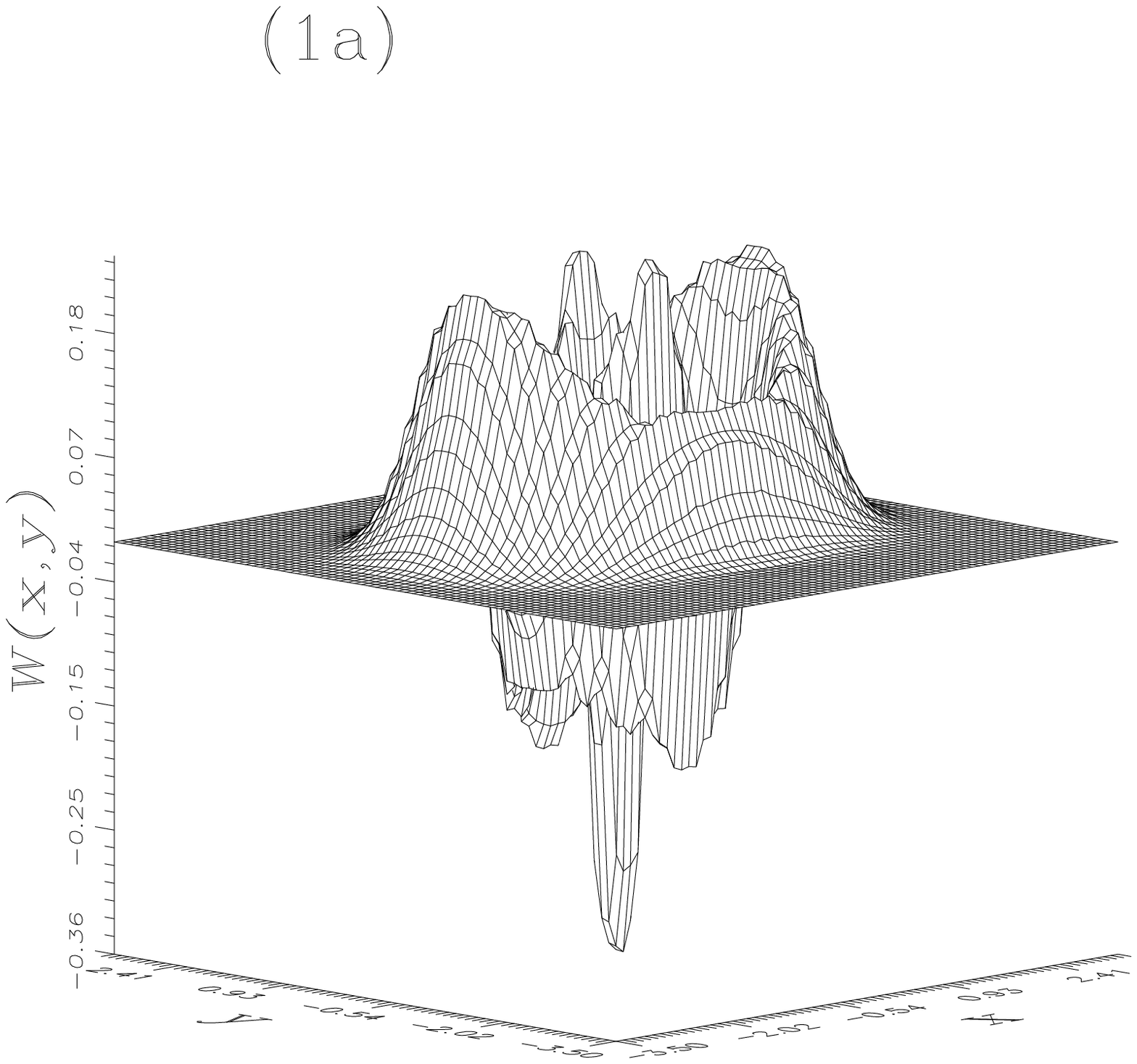}}
 \subfigure[]{\includegraphics[width=8cm]{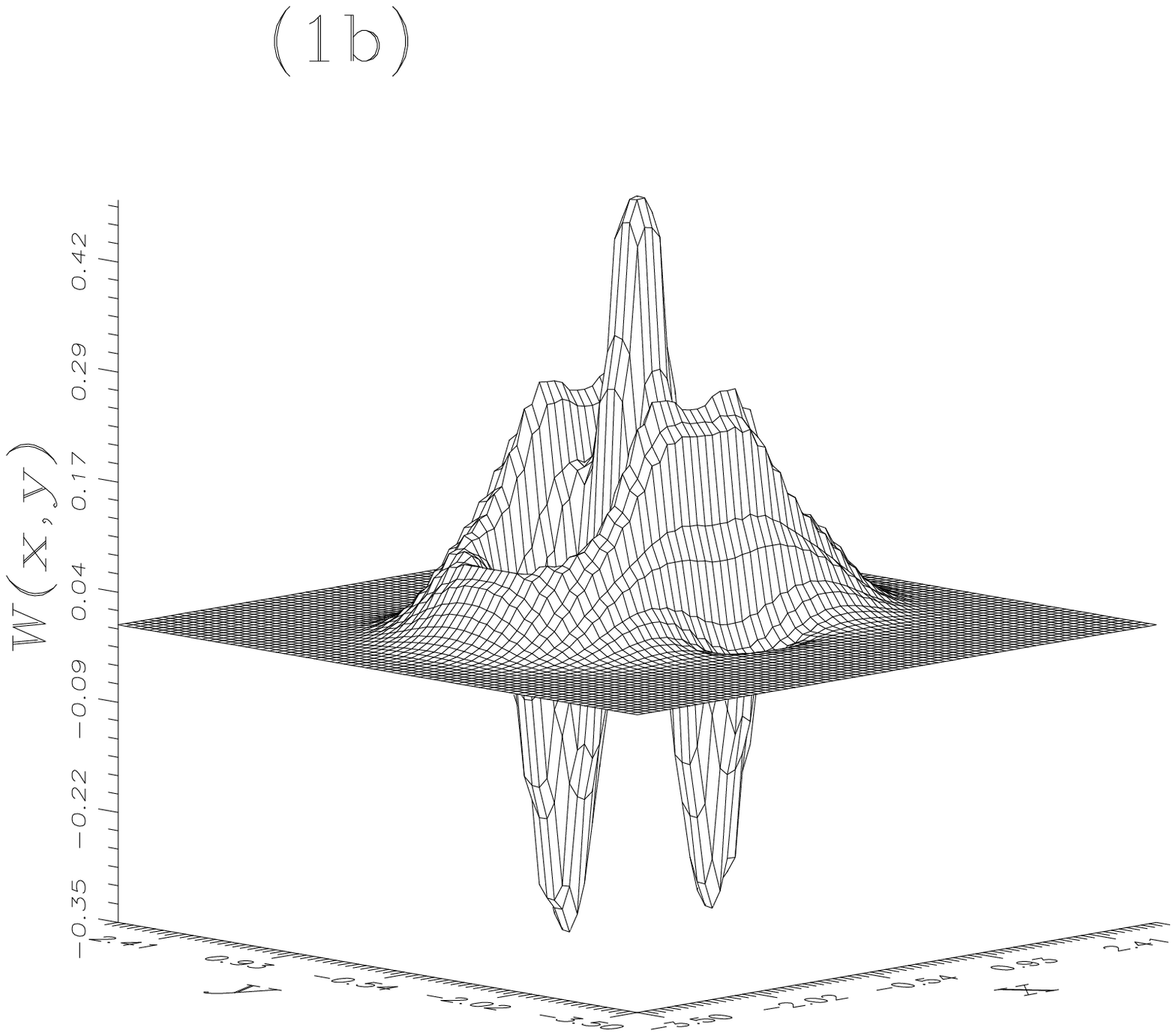}}
    \caption{
$W$-function against $x$ and $y$ for $n =1,\alpha =1$ and for a)
$\epsilon=1$; b) $\epsilon =-1$.  }
  \label{fig1}
\end{figure}

We start our discussion by focusing the
attention on the behaviour of the superposition of displaced number state.
It is important to mention that the properties of displaced number states
$\hat{D}(\alpha)|n\rangle$  have been
given in \cite{de} and their phase properties in \cite{tan}. In fact the
phase properties of such states are  interesting since they connect
the number state, which has no phase information,
and coherent states, which play the boundary role between the classical
and nonclassical
states and  always exhibit single peak structure of phase distribution.
The phase distribution of displaced number states exhibits nonclassical
oscillations with the number of peaks which are equal to $n+1$. This result is
interpreted in terms of the area of overlap in phase space \cite{tan}.
Unfortunately, the situation is completely different for the superposition
of displaced number state (see Fig. 2a and b for shown values of the
parameters).
From Fig. 2a it
is clear that the quantum interference between component states
$\hat{D}(\alpha)|n\rangle$ and $\hat{D}(-\alpha)|n\rangle$ leads to the
lateral nonclassical oscillations which become more pronounced and narrower
as $n$  increases. However, the statistical mixture of displaced number
states shows $n+1$ peaks around
 $\Theta =0$. We have not presented  the phase distribution for
 the odd-case since it is similar to that of the even-case.
 Comparing this behoviour with that of even (or odd) coherent states,
we conclude that  the  number states in the superposition make
the phase information more significant \cite{scr}. Turning our attention
to the Fig. 2b where $\alpha=1$, we can see that the distribution
of displaced number states exhibits two-peak structure as expected for
$n=1$ \cite{tan}, however, the distribution of even-case displays one
central peak
at $\Theta=0$ and two wings as $\Theta\rightarrow \pm \pi$, and finally
the distribution of odd-case provides four peaks. That is the
distribution here is
irregular, however, more smoothed than before and the structure of
the statistical
mixture (central part) is modified by the action of the interference term
arising from  the superposition of the states; this was the case for
$W$-function.

\begin{figure}[h]%
  \centering
  \subfigure[]{\includegraphics[width=8cm]{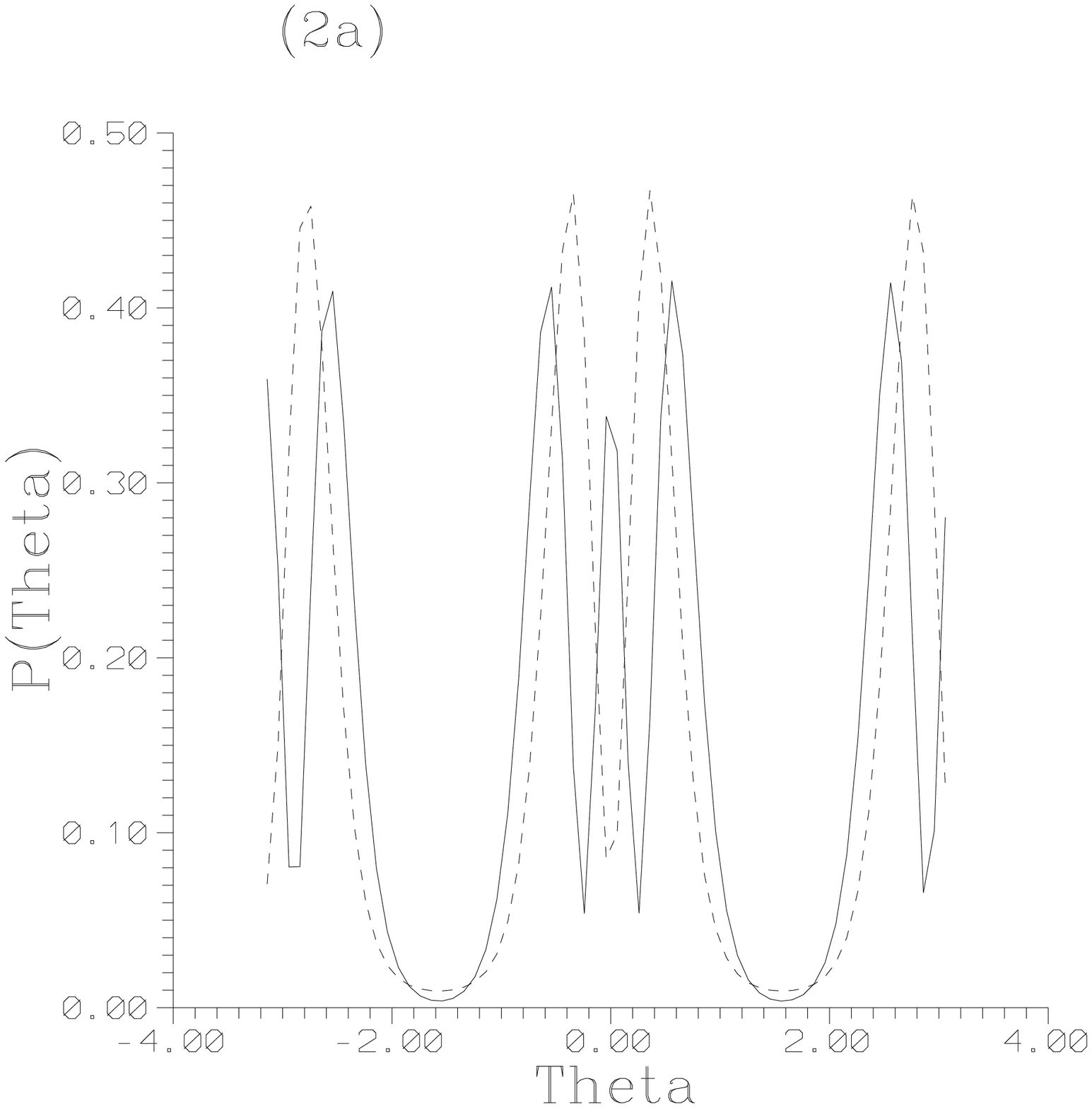}}
 \subfigure[]{\includegraphics[width=8cm]{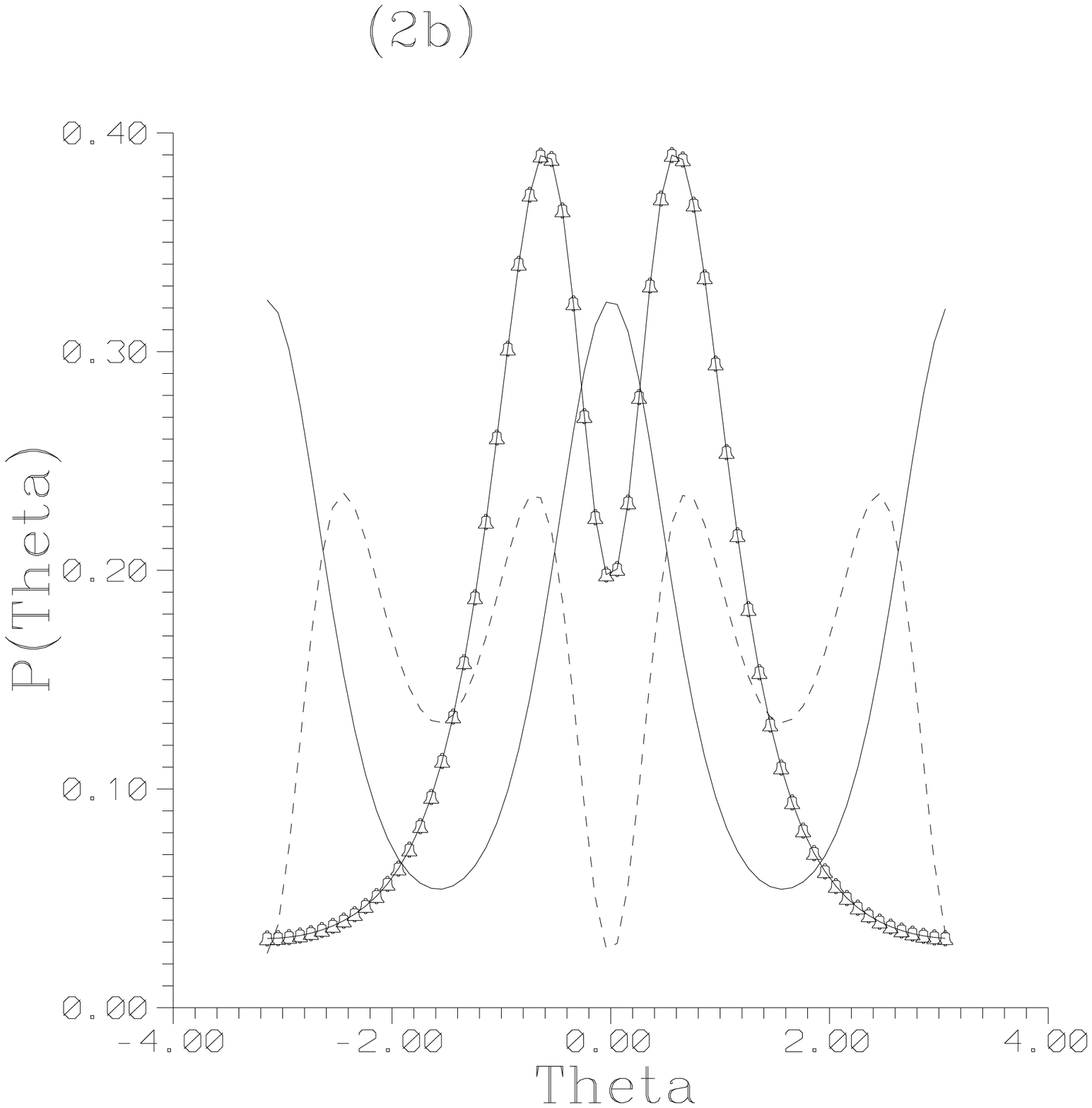}}
    \caption{
Phase distribution $P(\Theta)$ for superposition of displaced
number states and for a) $\alpha=2,\epsilon =1$ , $n=1$ (dashed
curve), and $2$ (solid curve); b) $ \alpha=1,n=1$ and $\epsilon
=0$ (bell-centered curve), $1$ (solid curve), and $-1$ (dashed
curve).  }
  \label{fig2}
\end{figure}

Before discussing the phase properties of the displaced and squeezed
number states it is reasonable to  remind the
behaviour of the well known squeezed states. As known for squeezed states
with non-zero displacement coherent amplitude, the phase distribution exhibits
the bifurcation phenomenon. In this phenomenon the single peak structure of
the coherent component is evolved into two peaks structure with respect to
both $\alpha $ (for large fixed value of squeezing parameter $r$) and $r$
(when $\alpha $ takes fixed value) \cite{schl1}. This phenomenon has been
recognized as a result of the competition between the two peaks structure of
the squeezed vacuum state and the single-peak structure of the coherent state.
For squeezed and displaced number states such phenomenon cannot occur
due to the effect of the Fock state which replaces the initial peak ($r=0$)
for coherent state by a multi-peak structure, i.e. by $n+1$ peaks (see Fig. 3a
for shown values of the parameters).   From this figure one can observe that
there is a three-peak structure corresponding to the case $n=2$ of
displaced
Fock state. The height of the central peak (i.e. at $\Theta=0$) is almost
the same and equals $(1/2\pi)|\sum_{m=0}^{\infty}C_{m}|^{2}$. That is
the central value of the phase distribution $P(\Theta=0)$ is insensitive
to squeezing provided that $r$ is finite.  However, the lateral  peaks
undergo phase squeezing as $r$ increases, i.e. the peaks become narrower.

Now  we can investigate  the behaviour of the superposition of
displaced and squeezed number states (see Figs. 3b-d for shown
parameters). Figs. 3b and 3c are given for the first regime for
even- and odd-cases, respectively. From Fig. 3b  we can see that
the initial oscillations are increased compared with Fig. 3a as a
result of the interference in phase space. Further, the initial
lateral peaks can evolve in the course of increasing $r$ to
provide bifurcation shape, i.e. the distribution curve undergoes a
transition from single- to a double-peaked form with increasing
$r$; however, the central peak is almost unchanged with increasing
$r$. This peak splitting is connected with the squeezed states
\cite{schl1}. Further, as the number of peaks  increases for
$r>0$, the distribution becomes more and more narrower. On the
other hand, the phase distribution of the odd-case is quite
different as we have shown before where the initial peaks are not
significantly changing as in the even-case. More precisely,
initial distribution becomes broader for a while and suddenly (at
$r\simeq 0.5$) breaks off to start to be a narrower distribution
for later $r$. It is clear that  in this regime the state
(\ref{1}) becomes more and more nonclassical \cite{fas} and  the
even- and odd-cases are distinguishable.
 Further, the phase distribution of the even-case is
more sensitive with respect to squeezing than the odd-case.
Nevertheless, for the second regime $\alpha>>1$ we noted that the phase
distribution carries
 at least the same initial information regardless  of the value of $r$
(see Fig. 3d for the shown values of the parameters).

\begin{figure}[h]%
  \centering
  \subfigure[]{\includegraphics[width=8cm]{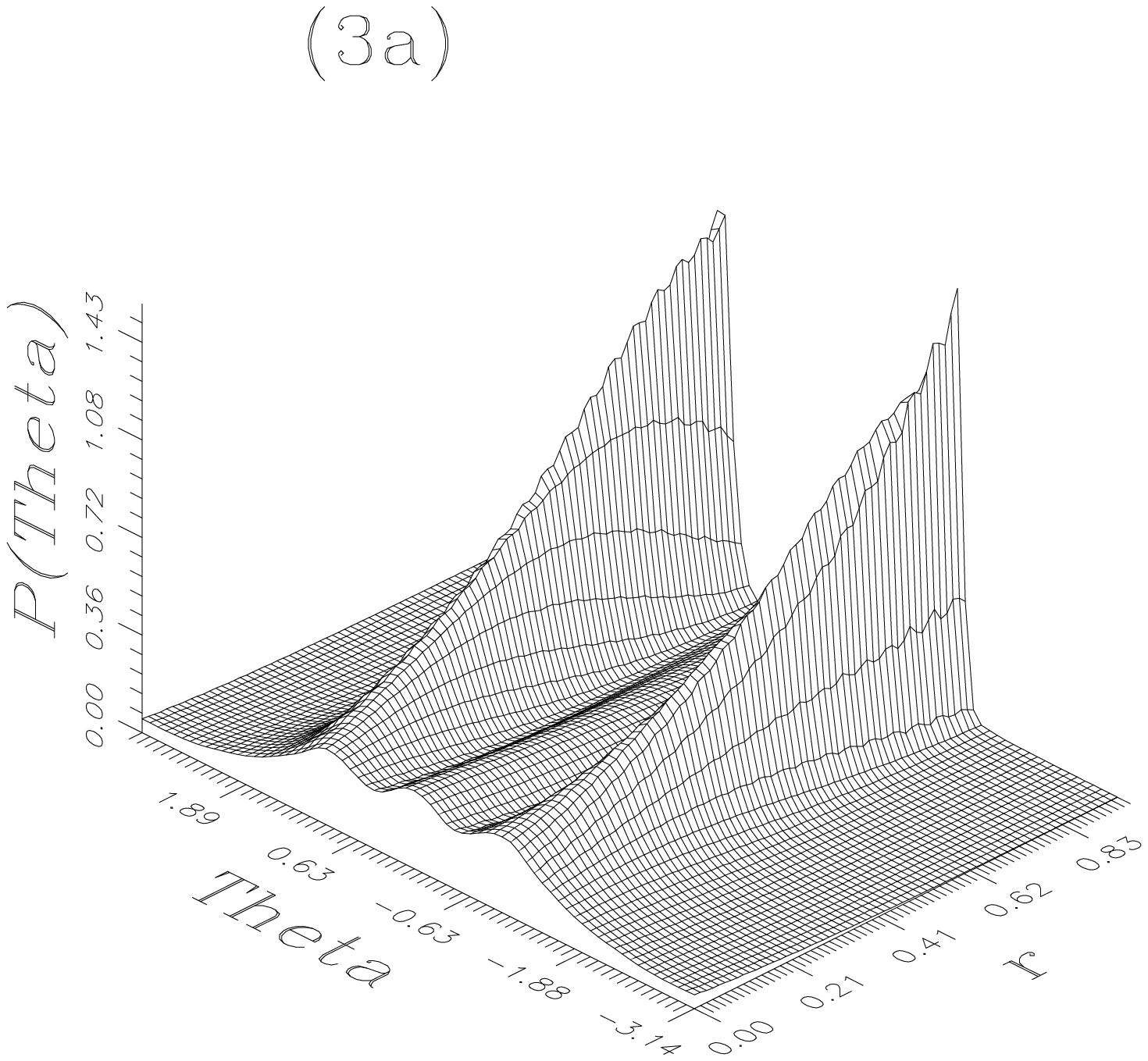}}
 \subfigure[]{\includegraphics[width=8cm]{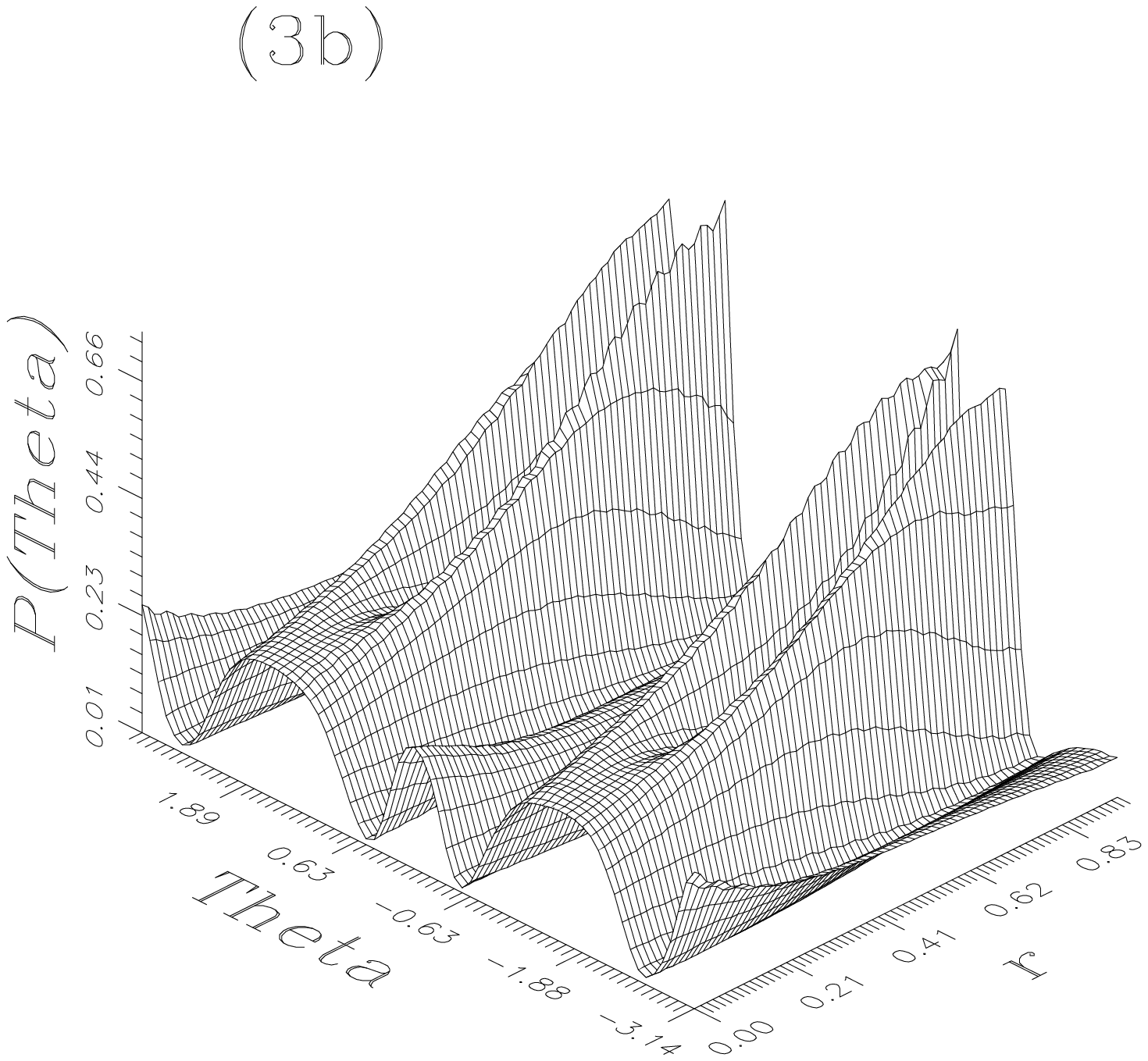}}
 \subfigure[]{\includegraphics[width=8cm]{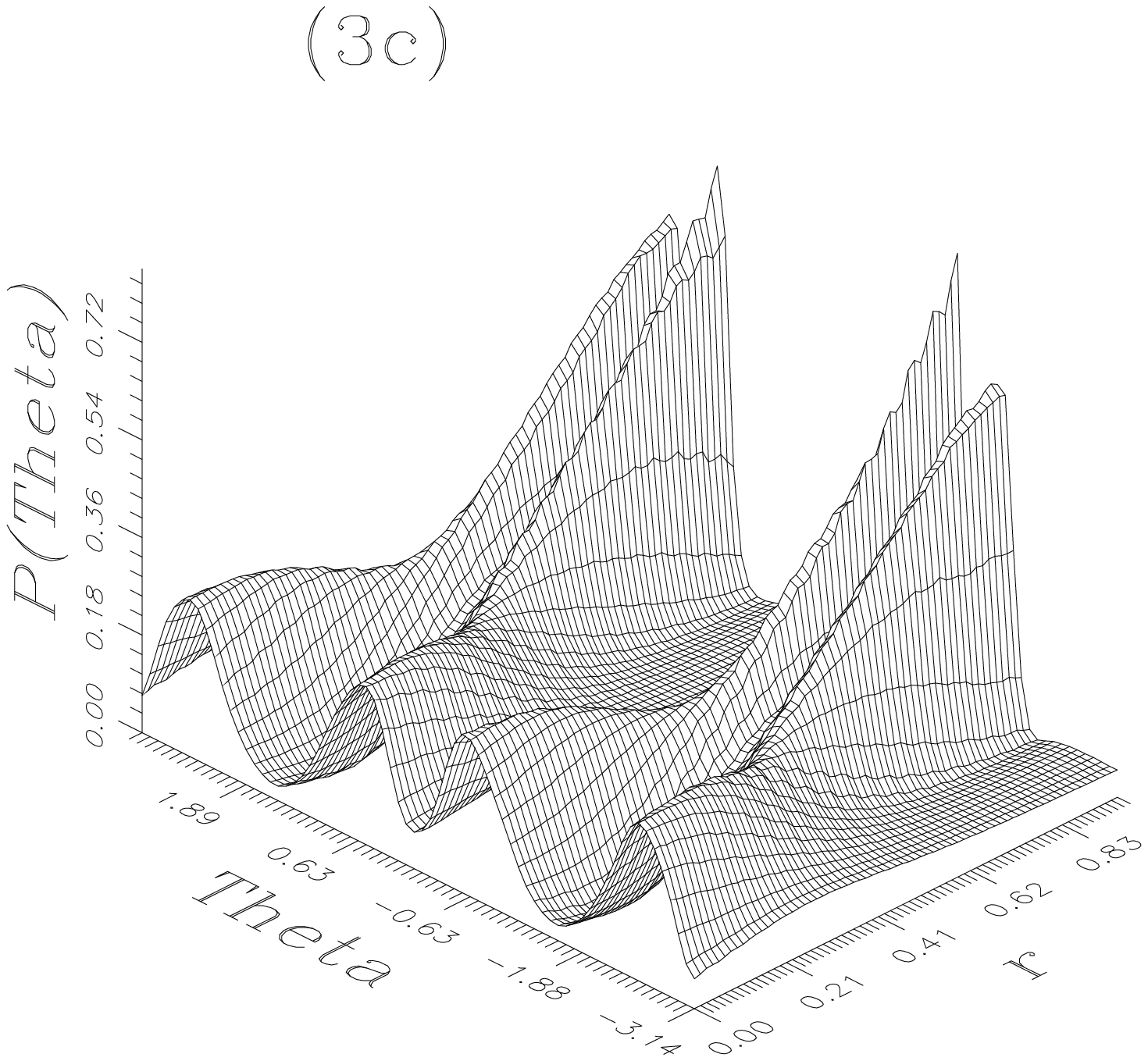}}
 \subfigure[]{\includegraphics[width=8cm]{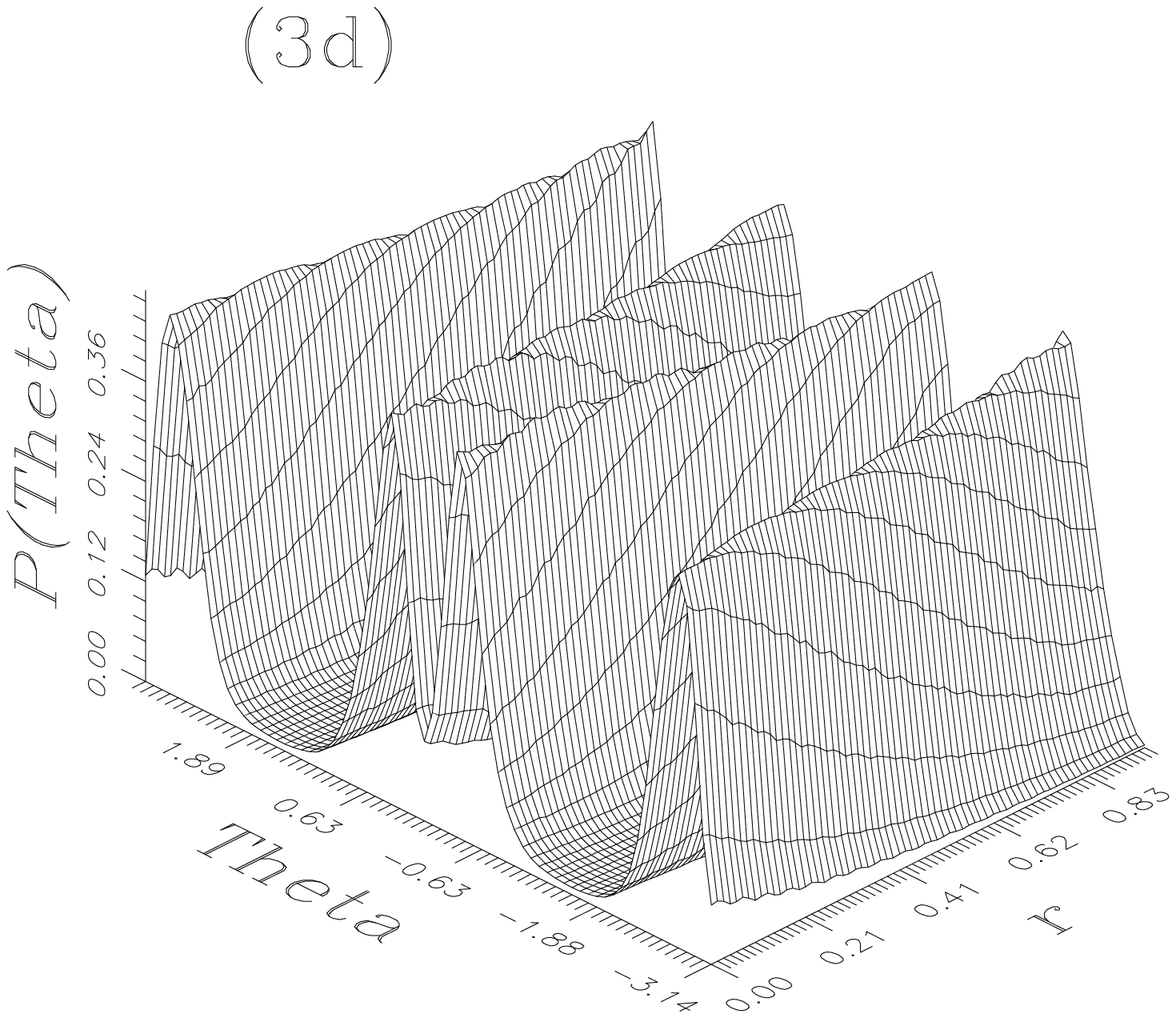}}
    \caption{
Phase distribution $P(\Theta)$ for superposition of displaced and
squeezed number states for $
(\epsilon,\alpha,n)=(0,1,2),(1,1,2),(-1,1,2)$ and $(1,2,2)$
corresponding to the cases a,b,c and d, respectively. }
  \label{fig3}
\end{figure}

\subsection{Variance, amplitude and phase squeezing}
In this subsection we investigate the behaviour of
the phase variance, and amplitude and phase fluctuations
following (\ref{20}), (\ref{15}) and (\ref{16}), respectively.
 We start our discussion by analysing the behaviour of
the superposition of displaced number states. For this purpose Figs. 4a and
4b are  shown for the phase variance, and amplitude and phase
fluctuations,
respectively. It is seen that  in general the phase
variance starts from the value $\pi^2/3$ (the vacuum state value)
and returns back  to it when $\alpha$ is large, but through different routes.
To be more specific, the phase variance of displaced number states starts
from the value for the vacuum, goes to a minimum, and then comes again
to $\pi^2/3$. However, the behaviour of  the superposition of
displaced number states takes different ways to arrive at the same result,
i.e. it starts from $\pi^2/3$ as before, goes to the
maximum value and eventually comes back to  the value of vacuum.
The comparison of the two cases shows the role
 of the quantum mechanical interference between state components.
Further, as $n$ increases, the oscillations in the variance  become
more pronounced.
Comparison of the behaviour of even- and odd-cases shows that
they are different
only over  the initial short interval of $\alpha$, i.e. when $\alpha$ is
small, and this agrees with what we have discussed earlier. So we can
conclude that for intensities high enough  of the coherent field, the
variance of the phase is approximately randomized. The route to this
randomization is dependent on the choice of $\epsilon$.
With respect to the amplitude and phase fluctuations, we can note
from (\ref{14}) that these quantities depend not only on the intensity of
the field, but also on the choice of the reference angle $\Theta_{0}$.
We have chosen here $\Theta_{0}=-\pi$, where the mean value  $|\langle
[\hat{N},\hat{\Phi}_{\theta}]\rangle|$ of squeezed displaced number
state approaches unity. Fig. 4b has been obtained to illustrate
 the parameters $S_{N}$ and $S_{\theta}$ which provide information
about the degree of squeezing in $\hat{n}$ and $\hat{\Phi}_{\theta}$.
One can observe from this figure that when $\epsilon =0$ (displaced number
state) and $\alpha\rightarrow 0$, the parameter $S_{N}$ tends to $-1$,
which means that the number state is $100\%$ squeezed with respect
to the operator $\hat{n}$. This situation is
expected since $\langle (\triangle \hat{n})^{2}\rangle=0$ for the number state.
Further, the larger the number of quanta is
the shorter is the interval  over which $S_{N}$
is squeezed. Also when $\alpha>>1$, squeezing in  $S_{\theta}$ is
remarkable, whereas $S_{N}$ becomes unsqueezed.
This result can be deduced from the behaviour of the phase variance
(see Fig. 4a), where $\langle (\triangle \hat{\Phi})^{2}\rangle\simeq 0$ at
$\alpha\simeq 4$ and this should be connected with maximum squeezing in
$S_{\theta}$ at this point.
Such behaviour of $S_{N}$ and $S_{\theta}$ confirms the fact that
the number of photons and phase are conjugate quantities in this approach.
As is known displaced
number states are not minimum uncertainty states and  the variances for
the quadrature operators never go below the standard quantum limit.
Moreover, they may exhibit sub-Poissonian statistics for the range
$\alpha^{2}\leq 1/2$ \cite{buz1}.
However, there is no relation between the sub-Poissonian statistics and
the fluctuation in the amplitude or the phase. This fact has been shown before
for the down-conversion process with quantum pump where the signal mode
can exhibit amplitude squeezing and at the same time it is
super-Poissonian \cite{tann}.

We proceed by discussing the behaviour of the superposition states
(long-dashed and circle-centered curves) in Fig. 4b where the interference
 in phase space starts
to play a role. We noted (from our numerical analysis) that only
the even case can provide squeezing in $S_{N}$ with maximum value at the
origin and squeezing interval larger than that discussed before. Indeed,
this maximum value
is related also to that of number state at $\alpha=0$.
It should be stressed here that $\alpha=0$ for the odd-case may lead to
a singularity.
As we mentioned before the superposition of displaced number states
 can exhibit strong sub-Poissonian character as well as
quadrature squeezing \cite{oba}.
Now we can illustrate the role of the squeezing in the superimposed
displaced number states optical cavity with respect to variance,
amplitude and
phase fluctuations. It is obvious, when squeezing is considered, that
the initial
value (at $\alpha=0$) of the phase variance is shifted  since we
have initially  squeezed number state which is providing phase information.
However, when $\alpha$ is large and $r$ is finite or also  $r$ is large,
it can be proved simply that the coefficient $C_{m}(r,\alpha,n,\epsilon)$
vanishes  and consequently the phase variance tends to $\pi^{2}/3$
(becomes randomized). Moreover,  the routs are here similar to those of
Fig. 4a. On the other hand,
squeezing could be seen
only in $S_{\theta}$ for  $\epsilon=0$ (see squared-centered curve in
Fig. 4b). Furthermore, comparison of the short-bell-centered curve (of
displaced Fock state) and squared-centered
curve reveals that squeezing parameter reduces the amount of squeezing in
$S_{\theta}$, too.
This means that the superposition of
displaced and squeezed number states which provide quadratures
squeezing has less information about amplitude and phase fluctuations.

\begin{figure}[h]%
  \centering
  \subfigure[]{\includegraphics[width=8cm]{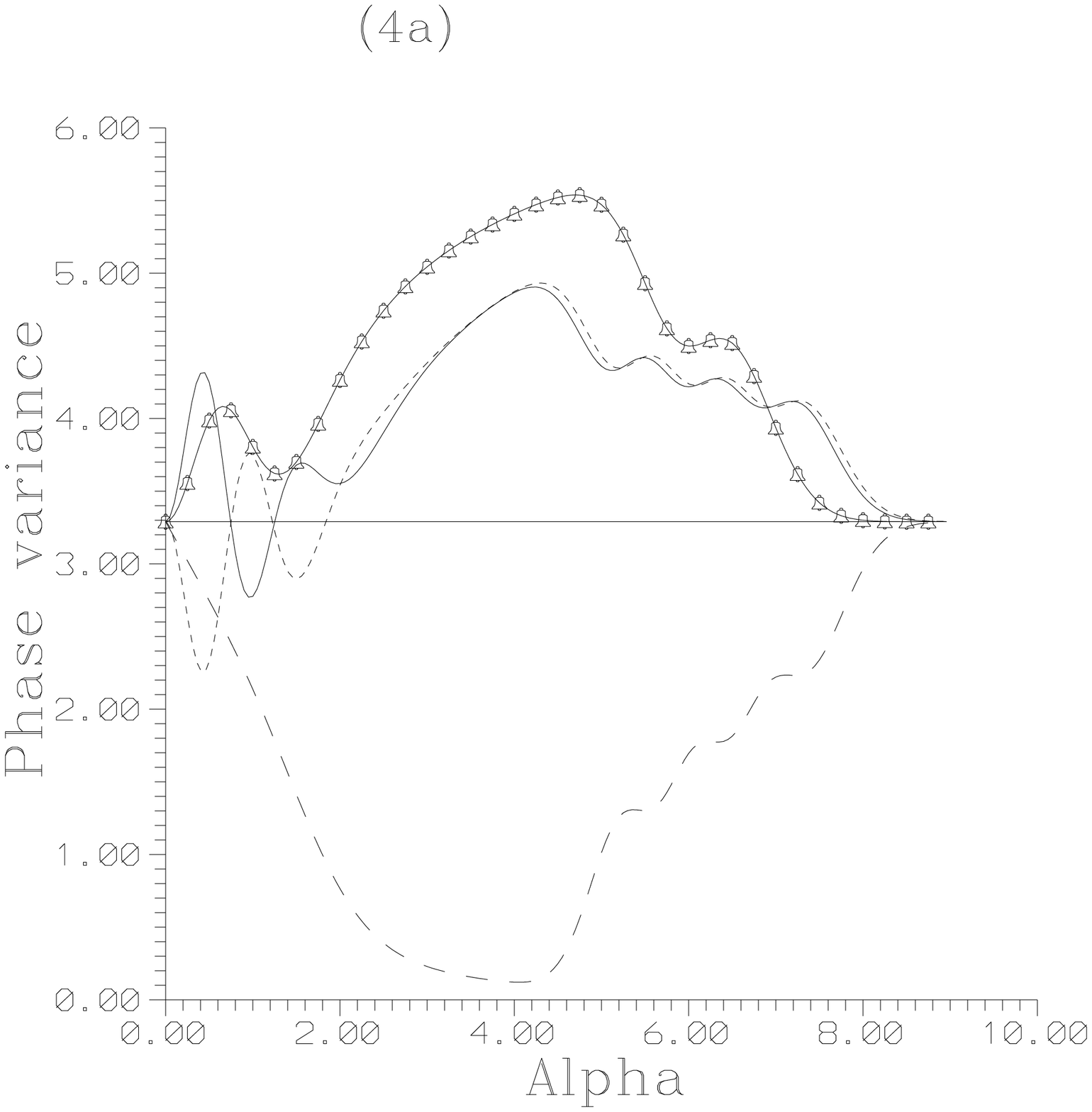}}
 \subfigure[]{\includegraphics[width=8cm]{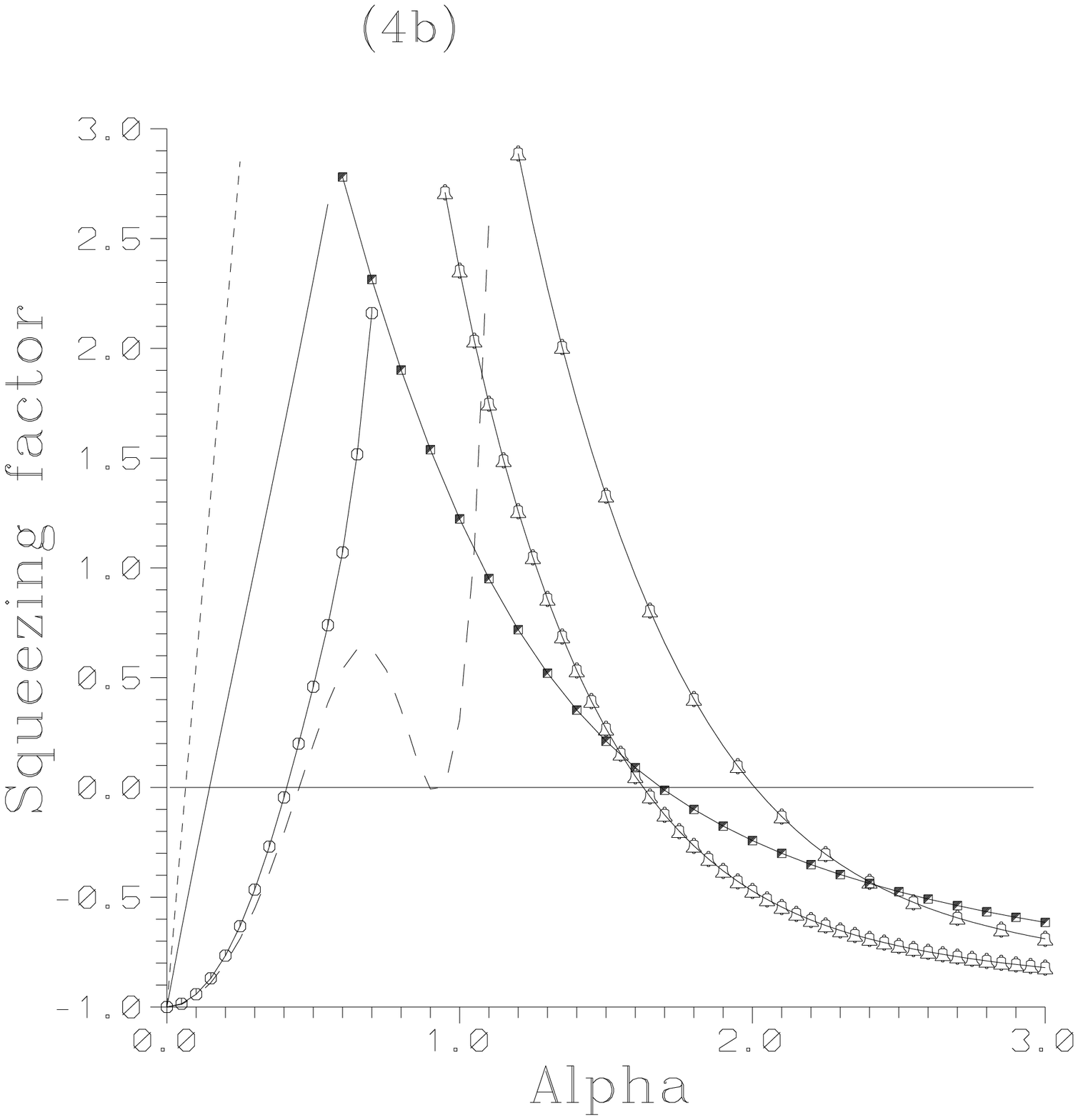}}
     \caption{
a) Phase variance of a superposition of displaced number states
against $\alpha$ for $(\epsilon,n)=(0,3)$ (long-dashed curve),
$(1,1)$ (bell-centered  curve), $(1,3)$ (solid  curve) and
$(-1,3)$ (short-dashed curve). The solid straight line is
corresponding to the phase variance of vacuum. b) Amplitude and
phase fluctuations of a superposition of displaced and squeezed
number states against $\alpha$ for $(\epsilon,n,r)=(0,1,0)$ (
$S_{N}$ solid curve and  $S_{\theta}$ short-bell-centered curve),
$(0,2,0)$
 ( $S_{N}$ short-dashed curve and  $S_{\theta}$ long-bell-centered curve),
$(1,1,0)$ ($S_{N}$  long-dashed curve), $(1,2,0)$ ($S_{N}$
circle-centered curve) and $(1,2,0.5)$ ($S_{\theta}$
squared-centered curve). The solid straight line is the bound of
squeezing. }
  \label{fig4}
\end{figure}

\section{Conclusions}
We have discussed the phase properties of the superposition of squeezed
and displaced number states from the point of view of the Pegg-Barnett
Hermitian phase formalism.
Moreover, the phase variance, the photon number and  phase
fluctuations have been discussed, too. The results have been explored
graphically.

We have shown that there are two regimes controlling
the behaviour of the phase for such superposition depending on whether
the superposition is macroscopic ($\alpha>>1$) or microscopic
($\alpha \leq 1$). This fact has been illuminated by means of the Wigner
function.
In the first regime,  the even- and odd-cases (i.e. $\epsilon=1$ and $-1$)
give similar behaviours and   the structure of the density matrix is
remarkable in figures.
All these facts are washed out
in the second regime where the behaviour  becomes irregular, however,
smooth.  In general we noted  that the higher the number of quanta is,
the more peaks the distribution possesses. Influence of the squeezing parameter
could be recognized in the second regime where the distribution exhibits
peak-splitting and peak-narrowing and this is in contrast with the fist
regime.

For  the phase variance we conclude that it asymptotically goes to
the value $\pi^2/3$ of the uniform distribution when either $\alpha$ or
$r$ is large, but through different routes. We have shown also that
this superposition can exhibit photon number fluctuations and phase
fluctuations.

{\bf Acknowledgments}

J. P. and F. A. A. E-O. aknowledge the partial support from the Project
VS96028 and Research Project CEZ: J 14/98 of Czech Ministry of Education
and from the Project 202/00/0142 of Czech Grant Agency.
One of us (M. S. A.) is grateful for the financial support
from the Project Math 1418/19 of the Research Centre, College of
Science, King Saud University.

\end{document}